\documentclass[12pt]{iopart}
\usepackage[dvips]{graphicx}
\usepackage{latexsym}
\usepackage{iopams}
\usepackage{amssymb}
\usepackage{verbatim,times,bbm}

\newcommand{\KK}{\mathcal{K}}
\newcommand{\lh}{\mathcal{L}_{\mathbbm{R}}(\mathcal{H})}
\newcommand{\elh}{\mathcal{L}(\mathcal{H})}
\newcommand{\lk}{\mathcal{L}_{\mathbbm{R}}(\mathcal{\KK})}
\newcommand{\densk}{\mathcal{D}(\KK)}
\newcommand{\lha}{\mathcal{L}_{\mathbbm{R}}(\hilba)}
\newcommand{\lhb}{\mathcal{L}_{\mathbbm{R}}(\hilbb)}
\newcommand{\ta}{\hat{\mathcal{T}}_\sistaa}
\newcommand{\tb}{\hat{\mathcal{T}}_\sistbb}
\newcommand{\ee}{\hat{\mathcal{E}}}
\newcommand{\ea}{\hat{\mathcal{E}}_\sistaa}
\newcommand{\eb}{\hat{\mathcal{E}}_\sistbb}
\newcommand{\ttt}{\hat{\mathcal{T}}}
\newcommand{\scrho}{\mathfrak{S}_{\hat{\rho}}}
\newcommand{\orb}{\widetilde{\mathfrak{S}}_{\hat{\rho}}}
\newcommand{\oto}{\mathrm{O}\otimes\mathrm{O}\hspace{0.4mm}(\mathcal{H})}
\newcommand{\otot}{\underline{\mathrm{O}}\otimes\underline{\mathrm{O}}\hspace{0.4mm}(\mathcal{H})}
\newcommand{\ost}{\mathrm{O}\otimes\mathrm{O}_{\mathrm{st}}(\hat{\rho})}
\newcommand{\oinv}{\mathrm{O}\otimes\mathrm{O}_{\mathrm{inv}}(\hat{\rho})}
\newcommand{\pto}{\mathrm{PTO}\otimes\mathrm{PTO}\hspace{0.4mm}(\mathcal{H})}
\newcommand{\cpto}{\mathrm{CPTO}\otimes\mathrm{CPTO}\hspace{0.4mm}(\mathcal{H})}
\newcommand{\ptm}{\mathrm{PT}(\mathcal{H})}
\newcommand{\pts}{\mathrm{PT}_{\mathfrak{S}}(\mathcal{H})}
\newcommand{\ot}{\overline{\mathrm{O}}\hspace{0.4mm}(\mathcal{H})}

\newcommand{\ototc}{\overline{\mathrm{O}}\otimes\overline{\mathrm{O}}\hspace{0.4mm}(\mathcal{H})}

\newcommand{\ext}{{\hspace{-1.8mm}\mbox{\tiny $\mathrm{ext}$}}}
\newcommand{\sss}{{\mathfrak{S}_{\hat{\rho}}}^\ext}

\newcommand{\ja}{\hat{J}_\sistaa}
\newcommand{\jb}{\hat{J}_\sistbb}
\newcommand{\jja}{\hat{\mathcal{J}}_\sistaa}
\newcommand{\jjb}{\hat{\mathcal{J}}_\sistbb}

\newcommand{\sista}{{\hspace{0.3mm}\mbox{\tiny $\mathsf{A}$}}}
\newcommand{\sistb}{{\hspace{0.3mm}\mbox{\tiny $\mathsf{B}$}}}
\newcommand{\sistaa}{{\mbox{\tiny $\mathsf{A}$}}}
\newcommand{\sistbb}{{\mbox{\tiny $\mathsf{B}$}}}
\newcommand{\hilba}{{\mathcal{H}_{\mbox{\tiny $\mathsf{A}$}}}}
\newcommand{\hilbb}{{\mathcal{H}_{\mbox{\tiny $\mathsf{B}$}}}}
\newcommand{\na}{{N_{\mbox{\tiny $\mathsf{A}$}}}}
\newcommand{\nb}{{N_{\mbox{\tiny $\mathsf{B}$}}}}
\newcommand{\rea}{{\mbox{\tiny $\mathrm{R}$}}}

\newcommand{\dd}{{\mathsf{d}}}
\newcommand{\DD}{{\mathsf{D}}}

\newcommand{\RR}{{\mathsf{R}}}

\newcommand{\hrho}{{\hat{\rho}}}

\newcommand{\dens}{\mathcal{D}(\mathcal{H})}

\newcommand{\minor}{{\mbox{$\mathrm{M}$}}}

\newtheorem{definition}{Definition}
\newtheorem{proposition}{Proposition}

\newtheorem{lemma}{Lemma}
\newtheorem{theorem}{Theorem}

\newcommand{\beq}{\begin{equation}}
\newcommand{\eeq}{\end{equation}}

\newcommand{\barr}{\begin{eqnarray}}
\newcommand{\earr}{\end{eqnarray}}
\newcommand{\Ord}[1]{{\cal O}\left( #1\right)}

\def\id{{\hat{\mathbb{I}}}}

\begin{document}

\title{Bipartite quantum systems: on the realignment criterion and beyond}

\author{Cosmo Lupo$^{1,2}$, Paolo Aniello$^{1,3}$, Antonello Scardicchio$^{4,5}$}

\address{$^1$ Dipartimento di Scienze Fisiche dell'Universit\`a di Napoli ``Federico II'', and INFN -- Sezione
di Napoli, via Cintia I-80126 Napoli, Italy}

\address{$^2$
Research Center for Quantum Information, Slovak Academy of Sciences,
D\'ubravsk\'a cesta 9, 845 11 Bratislava, Slovakia}

\address{$^3$ Facolt\`a di Scienze Biotecnologiche, Universit\`a
di Napoli ``Federico II'', Napoli, Italy}

\address{$^4$ Princeton Center for Theoretical Physics and Department of Physics
Princeton University, Princeton, NJ 08544, USA}

\address{$^5$
MECENAS, Universit\`a di Napoli ``Federico II", via Mezzocannone 8,
I-80134 Napoli, Italy}

\eads{\mailto{lupo@na.infn.it}, \mailto{aniello@na.infn.it},
\mailto{ascardic@princeton.edu}}

\begin{abstract}
Inspired by the `computable cross norm' or `realignment' criterion,
we propose a new point of view about the characterization of the
states of bipartite quantum systems. We consider a Schmidt
decomposition of a bipartite density operator. The corresponding
Schmidt coefficients, or the associated symmetric polynomials, are
regarded as quantities that can be used to characterize bipartite
quantum states. In particular, starting from the realignment
criterion, a family of necessary conditions for the separability of
bipartite quantum states is derived. We conjecture that these
conditions, which are weaker than the parent criterion, can be
strengthened in such a way to obtain a new family of criteria that
are independent of the original one. This conjecture is supported by
numerical examples for the low dimensional cases. These ideas can be
applied to the study of quantum channels, leading to a relation
between the rate of contraction of a map and its ability to preserve
entanglement.
\end{abstract}


\section{\label{intro}Introduction}

The relation between the state of a composite quantum system as a
whole and the configuration of its parts is a very peculiar feature
of quantum theory. As recognized since the early stages of
development of the theory~\cite{Einstein,Schr1,Schr2}, this is a
consequence of the tensor product structure of the state space of a
composite quantum system. This feature of quantum mechanics has its
most evident manifestation in the fact that it allows the presence
of non-classical correlations, i.e.\ of {\it entanglement}, between
the subsystems of a composite system. Nowadays, we may say that
quantum entanglement is not only regarded as a key for the
interpretation of quantum mechanics or as a mere scientific
curiosity, but also as a fundamental resource for quantum
information, communication and computation
tasks~\cite{NiCh,Bouwmeester}. However, despite the great efforts
made by the scientific community in the past decades, there are
still several open issues regarding the mathematical
characterization of composite quantum states, even in the
`elementary case' of a bipartite system with a finite number of
levels.

A major challenge is to characterize those states of a bipartite
system that are entangled. According to the definition due to R.\
F.\ Werner~\cite{Werner}, \emph{entangled} (mixed) states differ
from \emph{separable} states since they cannot be prepared, not even
in principle, from product states by means of local operations and
classical communication only. In mathematical terms, a (mixed) state
$\hrho$ --- a positive (trace class) operator of unit trace
--- in a composite Hilbert space $\mathcal{H}=\hilba\otimes\hilbb$
is called \emph{separable} if it can be represented as a convex sum
of product states:
\begin{equation} \label{separable}
\hrho = \sum_i p_i\; \hrho_i^\sista \otimes \hrho_i^\sistb,
\end{equation}
with $p_i \geq 0$ and $\sum_i p_i = 1$; otherwise, $\hrho$ is said
to be \emph{entangled}. We remark that, if $\hrho$ is separable,
decomposition~{(\ref{separable})} is in general not unique, and the
smallest number of terms in the sum (usually called
\emph{cardinality}), due to Caratheodory's theorem, is not larger
than the squared dimension of the \emph{total} Hilbert space of the
system $\mathcal{H}$ (see~\cite{Horo}).

Since quantum entanglement is a very important subject, also in view
of its several potential applications, separability criteria are
regarded as extremely precious tools. Among a plethora of proposed
separability criteria --- i.e.\ suitable conditions satisfied by all
separable states whose violation allows to detect entanglement (see,
for instance, \cite{Horo96,3H,Horo99,Nielsen})
--- the present contribution is mainly inspired by the criterion that
was proposed in~\cite{ReCr} with the name of `realignment criterion'
(RC) and in~\cite{CCN} with the name of `computable cross norm'
criterion. As we will try to argue, the RC brings attention to the
role played by the \emph{Schmidt coefficients}~\cite{Schmidt} of a
bipartite quantum state in the characterization of entanglement.
Trying to shed light on this role will be the main goal of our
contribution.

The paper develops along the following lines.
In section~\ref{class:def}, we introduce a `Schmidt equivalence
relation' in the set of states of a bipartite quantum system, and we
show the link between this notion and the RC.
Section~\ref{class:study} is devoted to the characterization of the
Schmidt equivalence classes. We follow two different approaches: the
characterization of some groups acting on the Schmidt equivalence
classes and the analysis of the local geometry of these equivalence
classes regarded as manifolds.
A family of separability criteria is presented in
section~\ref{ideas}, which are extensions of the RC. These criteria
are based on the `symmetric polynomials' in the Schmidt
coefficients, and are weaker than the parent criterion.
In section~\ref{duality}, the well known correspondence between
quantum states and quantum maps (see, e.g., \cite{MaSt,Jam,ZycBen}),
i.e.\ completely positive trace-preserving (CPT) maps, is
considered, and a straightforward application of the derived family
of separability criteria to the study of CPT maps is discussed.
Through this correspondence, separable states are associated to
entanglement breaking (EB) channels~\cite{Holevo}. The proposed
family of criteria, applied to this context, leads to a purely
geometrical characterization of EB maps.
In section~\ref{new}, we formulate the conjecture that the proposed
criteria can be strengthened in order to obtain new necessary
conditions for separability which are independent of the parent RC.
Numerical examples in support of this thesis are provided in
section~\ref{examples} for low dimensional bipartite systems.

\section{\label{class:def} Schmidt equivalence classes of states of a bipartite quantum system}

Let us consider a bipartite, finite-dimensional, complex Hilbert
space $\mathcal{H}=\hilba\otimes\hilbb$ ---
$\hilba\cong\mathbbm{C}^\na$, $\hilbb\cong\mathbbm{C}^\nb$,
$\na,\nb\ge 2$
--- and the corresponding real vector spaces of Hermitian operators
$\lh$, $\lha$, and $\lhb$ (the spaces of observables) in
$\mathcal{H}$, $\hilba$ and $\hilbb$, respectively, that are
naturally endowed with a scalar product, namely, the \emph{bilinear}
Hilbert-Schmidt (HS) product:
\begin{equation}
\langle \hat{A} , \hat{B} \rangle_\mathrm{HS} = \tr( \hat{A}
\hat{B}),\ \ \ \mbox{($\hat{A}$, $\hat{B}$ Hermitian)}.
\end{equation}
In particular, the \emph{density operators} --- i.e.\ the positive
operators of unit trace --- in $\mathcal{H}$ ($\hilba$ and $\hilbb$,
respectively) can be regarded as elements of the real vector space
$\lh$ ($\lha$ and $\lhb$, respectively), in which they form a convex
body that will be denoted by $\dens$. Observe that
$\lh=\lha\otimes\lhb$, with $\lha\cong\mathbbm{R}^{\na^2}$ and
$\lhb\cong\mathbbm{R}^{\nb^2}$. The HS product allows to write a
(nonunique) \emph{Schmidt decomposition}~\cite{Schmidt} of a density
operator $\hrho\in\dens$, i.e.\footnote{We remark that actually any
operator in $\lh$ admits a Schmidt decomposition.}
\begin{equation}\label{state}
\hrho = \sum_{k=1}^\dd \lambda_k\, \hat{F}_k^\sista \otimes
\hat{F}_k^\sistb,\ \ \ \lambda_1\ge\lambda_2\ge\ldots\ge
\lambda_\dd\ge 0,
\end{equation}
where
\begin{equation}
\tr({\hat{F}_h^\sista} \hat{F}_k^\sista) =\delta_{hk}=
\tr(\hat{F}_h^\sistb \hat{F}_k^\sistb),\ \ \ h,k\in\{1,\ldots\dd\},
\end{equation}
and the real positive numbers $\{\lambda_k\}_{k=1}^\dd$ are the
(uniquely determined) \emph{Schmidt coefficients} (in short, SC's).
Note that the number of terms in the sum equals $\dd =
\min\{\na^2,\nb^2\}$ (we will also set $\DD=\max\{\na^2,\nb^2\}$).
The definition of the SC's of a bipartite density operator is the
natural generalization of the standard definition for pure states
(see, for instance, \cite{Aniello}). It is worth stressing that,
since the operators forming the orthonormal systems
$\{\hat{F}_k^\sista\}_{k=1}^\dd$, $\{\hat{F}_k^\sistb\}_{k=1}^\dd$
(in $\lha$ and $\lhb$, respectively) are Hermitian, the operators
$\{\hat{F}_k^\sista \otimes \hat{F}_k^\sistb\}_{k=1}^\dd$ are
\emph{observables}. Hence
--- at least in principle --- \emph{the SC's are physically measurable quantities}:
\begin{equation}
\lambda_k = \tr( \hrho\, \hat{F}_k^\sista \otimes \hat{F}_k^\sistb).
\end{equation}
The set of `local' operators $\{\hat{F}_k^\sista \otimes
\hat{F}_k^\sistb\}_{k=1}^\dd$ are also referred to as {\it local
orthogonal observables} \cite{LOO}. Decomposition~{(\ref{state})}
has been recently considered --- see~\cite{arrange} --- in
connection with the formulation of new separability criteria.

We observe that the convex body $\dens$ can also be regarded as
immersed in the \emph{complex} vector space $\elh$ of linear
operators in $\mathcal{H}$, vector space that can be endowed with
the \emph{sesquilinear} HS product (denoted, again, as $\langle
\cdot , \cdot \rangle_\mathrm{HS}$). Then, one can consider a
Schmidt decomposition of a density operator $\hrho$ in $\mathcal{H}$
with respect to the complex Hilbert space $\elh$. It is clear that
such a decomposition will contain the same SC's as
decomposition~{(\ref{state})}, but this time will involve an
orthonormal system of, in general, non-Hermitian operators.

Given a bipartite density operator, one can uniquely determine its
SC's. On the other hand, it is clear that the SC's do not identify a
unique quantum state. It is then natural to formulate the following
definition:
\begin{definition}[Schmidt equivalence relation] \label{Sch-equiv}
We say that two bipartite density operators are Schmidt equivalent
if they share the same set of Schmidt coefficients.
\end{definition}
Then the convex set of density operator in the Hilbert space
$\mathcal{H}$ --- which will be denoted by $\dens$ --- is
partitioned into Schmidt equivalence classes.

It is known that a bipartite {\it pure} state is completely
characterized, with respect to entanglement, by the corresponding
SC's~\cite{more_than_one}. Although the same characterization cannot
be extended to a generic state, it is reasonable to suppose that
there exists some relation between the SC's of a bipartite density
operator and the entanglement properties of this state, and to
address the following questions: {\it``What relevant properties are
encoded by the SC's of a bipartite density operator, and how one can
characterize the Schmidt equivalence classes?"} In the following, we
will try to analyze these questions and to provide some reasonable
answers.

A first observation is that the SC's determine the \emph{purity} of
a state. Let us recall that the purity is defined as the trace of
the square of the density operator:
\begin{equation}
\mathcal{P}(\hrho) := \tr\left(\hrho^2\right);
\end{equation}
hence, $\mathcal{P}(\hrho)\in ]0,1]$, $\forall\hrho\in\dens$. It
follows from the definition of Schmidt decomposition that the purity
equals the sum of the squares of the SC's:
\begin{equation}\label{purity}
\frac{1}{\na\nb} \le \mathcal{P}(\hrho) =\langle \hrho , \hrho
\rangle_\mathrm{HS}= \sum_{k=1}^\dd \lambda_k^2\le 1.
\end{equation}
Thus, the purity is the simplest property which is (completely)
described by the SC's of a density operator. Another relevant fact
is the existence of a link between the separability of a density
operator and its SC's.

Indeed, the `realignment criterion' (in short RC
see~\cite{ReCr,CCN}; see also~\cite{Aniello}, where a generalization
of the RC is obtained) establishes a necessary condition for the
separability of a quantum state (or a sufficient condition for the
nonseparability). It can be formulated in various equivalent ways.
From our point of view, it can be regarded as a condition on the
SC's of a separable density operator $\hrho$. Precisely, it imposes
an upper bound for the sum of its SC's:
\begin{theorem}[the `realignment criterion']
If a bipartite density operator $\hrho$ is separable, then its
Schmidt coefficients $\{\lambda_k\}_{k=1}^\dd$ satisfy the following
inequality:
\begin{equation}\label{rc_cond}
\sum_{k=1}^\dd \lambda_k \leq 1.
\end{equation}
\end{theorem}
The RC is easily implementable. In particular, in~\cite{ReCr} it has
been introduced the notion of a {\it realigned matrix} $\rho^\rea$
associated with the bipartite density operator $\hrho$ in order to
compute the l.h.s.\ of inequality~{(\ref{rc_cond})}. Fixed
orthonormal bases $\{|n\rangle\}_{n=1,\dots\na}$ and
$\{|\nu\rangle\}_{\nu=1,\dots\nb}$ in the local Hilbert spaces
$\hilba$, $\hilbb$, respectively, and assuming that
$\rho_{(m\mu)(n\nu)}$ is the representative matrix of the density
operator $\hrho$ with respect to the product basis
$\{|n\rangle\otimes|\nu\rangle\}_{n=1,\dots\na}^{\nu=1,\dots\nb}$---
where $(m\mu)$, $(n\nu)$ are double indexes --- the corresponding
realigned matrix (with respect to the given basis) is defined as
\begin{equation}\label{rea_matrix}
\rho^\rea_{(mn)(\mu\nu)} := \rho_{(m\mu)(n\nu)}.
\end{equation}
It turns out that the SC's of $\hrho$ are the {\it singular values}
of the realigned matrix $\rho^\rea$, see~\cite{Aniello}. Precisely,
consider a \emph{singular value decomposition} of the realigned
matrix, i.e.
\begin{equation}\label{ref_me}
\rho^\rea = \mathcal{U} \Delta \mathcal{V},
\end{equation}
where $\mathcal{U}$ and $\mathcal{V}$ are unitary matrices,
belonging respectively to the unitary groups $\mathrm{U}(\na^2)$ and
$\mathrm{U}(\nb^2)$, and $\Delta$ is a rectangular matrix such that
its nonvanishing entries are positive and placed along the principal
diagonal only. Then, the diagonal entries of $\Delta$ are the
singular values of $\rho^\rea$, hence, the SC's of $\hrho$. At this
point, observing that the singular values of the realigned matrix
$\rho^\rea$ coincide with the eigenvalues of the positive matrix
$|\rho^\rea|=\sqrt{{\rho^\rea}^\dag\rho^\rea}$, one concludes that
inequality~{(\ref{rc_cond})} can also be written as
\begin{equation}
\|\rho^\rea\|_{\tr}=\tr\left(|\rho^\rea|\right)\leq 1.
\end{equation}

Given a density operator $\hrho\in\dens$, we will denote by $\scrho$
the Schmidt equivalence class containing $\hrho$. Beside the Schmidt
equivalence class $\scrho$, we will consider the \emph{extended
Schmidt equivalence class} containing $\hrho$, namely, the set
$\sss$ of all the Hermitian operators in $\mathcal{H}$ that share
with $\hrho$ the same set of Schmidt coefficients. Thus, we have
that $\scrho=\sss\hspace{-0.3mm}\cap\dens$.

\section{\label{class:study}Characterization of the Schmidt equivalence classes}

Aim of the present section is to give a basic characterization of
the Schmidt equivalence classes of states; see
Definition~{\ref{Sch-equiv}}. In this regard, one can adopt two
different approaches. On one hand, one can try to characterize some `natural'
groups for which an action on the Schmidt equivalence classes is
defined; see subsection~{\ref{local_maps}}. On the other hand, one can
regard the Schmidt equivalence classes as manifolds and study
their local geometry considering the action of the local orthogonal
groups; see subsection~{\ref{local-analysis}}. Although the results
that we obtain are still somewhat `preliminary', we think that it is
worthwhile to report them since a description of Schmidt equivalence
classes seems to be completely missing in the literature.

\subsection{Groups acting on the Schmidt equivalence classes} \label{local_maps}

Note that, since $\lh$, $\lha$ and $\lhb$ are \emph{real} Hilbert
spaces, the unitary (super)operators in these spaces belong to
orthogonal groups. For instance, a unitary operator in $\lh$ belongs
to the orthogonal group $\mathrm{O}(\na^2\nb^2)$. The class of
unitary operators in $\lh$ that are decomposable as the tensor
product of two unitary operators in $\lha$ and $\lhb$, respectively
--- i.e.\ of the form $\ta \otimes \tb$ , with $\ta$ in $\mathrm{O}(\na^2)$,
$\tb$ in $\mathrm{O}(\nb^2)$
--- will be denoted by $\oto$. It is clear that the maps in $\oto$
preserve the SC's of every element of $\lh$, but, in general,
$\ta\otimes\tb\,(\dens)\not\subset\dens$. Note, moreover, that the
set $\oto$ is a group (isomorphic to the direct product
$\mathrm{O}(\na^2)\times \mathrm{O}(\nb^2)$) with respect to the
usual composition of maps. The orbit in $\lh$, under the action of
this group, passing through $\hrho\in\dens$ will be denoted by
$\orb$; i.e.
\begin{equation}
\orb := \{\ttt (\hrho)\in\lh\colon\ \ttt=\ta \otimes \tb\in\oto\}.
\end{equation}
\begin{proposition} \label{equivalence}
The orbit $\orb$ of the group $\oto$ through $\hrho\in\dens$
coincides with the extended Schmidt equivalence class containing
$\hrho$:
\begin{equation}
\orb=\sss .
\end{equation}
It follows that
\begin{equation}
\scrho=\orb\cap\dens .
\end{equation}
Therefore, two states $\hat{\rho}$ and $\hat{\sigma}$ in $\dens$ are
Schmidt equivalent if and only if
\begin{equation}\label{equivalence_eq}
\hat{\rho} = \ta \otimes \tb\, (\hat{\sigma}),
\end{equation}
for some unitary operators $\ta$ and $\tb$ in $\lha$ and $\lhb$,
respectively.
\end{proposition}

\noindent {\bf Proof:} As already observed, given $\hrho\in\dens$,
for every $\ttt$ in the group $\oto$, $\ttt(\hrho)$ belongs to
$\sss$; hence, $\orb\subset\sss$. On the other hand, let $\hrho =
\sum_{k=1}^\dd \lambda_k\, \hat{F}_k^\sista \otimes
\hat{F}_k^\sistb$ be a Schmidt decomposition of $\hrho$ and
$\hat{C}= \sum_{k=1}^\dd \lambda_k\, \hat{G}_k^\sista \otimes
\hat{G}_k^\sistb$ a Schmidt decomposition of an arbitrary element
$\hat{C}$ of $\sss$. Then, for every couple of unitary operators
$\ta$ and $\tb$ in $\lha$ and $\lhb$, respectively, such that $\ta
(\hat{F}_k^\sista)=\hat{G}_k^\sista$ and $\tb
(\hat{F}_k^\sistb)=\hat{G}_k^\sistb$, $k=1,\ldots,\dd$, we have:
$\ta\otimes\tb (\hrho)=\hat{C}$. Hence,
$\orb\supset\sss$.~{$\square$}

The set of all the maps of the form $\ta \otimes \tb$ --- with $\ta$
and $\tb$ unitary operators in $\lha$ and $\lhb$, respectively ---
such that $\scrho$ is \emph{stable} under the action of $\ta \otimes
\tb$, i.e.\ such that
\begin{equation}
\ta \otimes \tb\,(\scrho)\subset\scrho,
\end{equation}
is a semigroup (with respect to composition) with identity,
contained in the group $\oto$, semigroup which will be denoted by
$\ost$. The subset $\oinv$ of $\ost$ defined by
\begin{equation}
\oinv :=\{\ta \otimes \tb\in\ost\colon (\ta \otimes
\tb)^\dagger\in\ost\}
\end{equation}
is a group. It is easy to check that $\oinv$ coincides with the
subset of $\ost$ containing those maps that leave $\scrho$
\emph{invariant}, i.e.
\begin{equation}
\oinv =\{\ta \otimes \tb\in\ost\colon \ta \otimes
\tb\,(\scrho)=\scrho\}.
\end{equation}
As an example of an operator belonging to $\oinv$, for all
$\hrho\in\dens$, consider the linear map
$\jja\otimes\jjb\colon\lh\rightarrow\lh$, with $\jja$ and $\jjb$
unitary operators defined by:
\begin{equation}
\hspace{-1.8cm}\jja\colon \lha\ni\hat{A}\mapsto
\ja\hat{A}\ja\in\lha, \ \ \jjb\colon \lhb\ni\hat{B}\mapsto
\jb\hat{B}\jb\in\lhb,
\end{equation}
where $\ja$ and $\jb$ are `local' \emph{complex conjugations} (i.e.\
selfadjoint antiunitary operators) in $\hilba$ and $\hilbb$,
respectively. The maps $\ja$ and $\jb$ are partial transpositions,
so that the map $\jja\otimes\jjb$ is the transposition associated
with a tensor product basis in $\hilba\otimes\hilbb$ (recall that
transposition, as complex conjugation, is a basis-dependent notion).
As it is well known, a transposition is a \emph{positive
trace-preserving} map (in short, PT map), and it is selfadjoint with
respect to the HS scalar product. Therefore, the selfadjoint unitary
operator $\jja\otimes\jjb$ is contained in the group $\oinv$, for
all $\hrho\in\dens$.

It is natural to wonder how states belonging to the same Schmidt
equivalence class can be connected by physically realizable
transformations. We will then consider the semigroup with identity
of PT maps in $\lh$, which will be denoted by $\ptm$. It is worth
defining the following subset of $\ptm$:
\begin{eqnarray}
\pts & := & \{\ee\in\ptm\colon\ \ee\, \mbox{bijective},\ \
\ee^{-1}\in \ptm,\nonumber\\
& &\ \ee(\scrho)=\scrho,\ \forall\hrho\in\dens\}.
\end{eqnarray}
It is clear that the set $\pts$ is a group.

As already observed, unitary maps in $\lh$ of the form
$\ta\otimes\tb$ --- with $\ta$ and $\tb$ unitary operators in $\lha$
and $\lhb$, respectively --- preserve the SC's of every element of
$\lh$, but, in general, $\ta\otimes\tb\,(\dens)\not\subset\dens$. It
is therefore natural to consider the class of linear maps in $\lh$
belonging to the set $\oto\cap\ptm$. It is clear that this set is a
semigroup (with respect to composition of maps) with identity
contained in $\ost$. We will show now that it is actually a group
which is a subgroup of $\oinv$. We need two preliminary results; the
proof of the first one is trivial.
\begin{lemma} \label{primo}
The inverse of a bijective trace-preserving map from $\lh$ onto
$\lh$ is trace-preserving.
\end{lemma}
\begin{lemma} \label{secondo}
A positive linear map $\ttt\colon\lh\rightarrow\lh$ which is unitary
transforms the convex cone of positive operators in $\mathcal{H}$
onto itself; therefore, $\ttt^\dagger$ is a positive map. Hence, in
particular, if a linear map belonging to $\oto$ is a positive map,
then its inverse is a positive map too.
\end{lemma}
\noindent {\bf Proof:} We will prove the statement by contradiction.
Suppose that $\hat{B}=\ttt(\hat{A})$ is positive, and assume that
$\hat{A}$ is not a positive operator. Then, for some
$\psi\in\mathcal{H}$, we should have that
$\langle\psi,\hat{A}\,\psi\rangle <0$; hence,
$\tr(\ttt(|\psi\rangle\langle\psi |)\,
\ttt(\hat{A}))=\tr(|\psi\rangle\langle\psi |\,
\hat{A})=\langle\psi,\hat{A}\,\psi\rangle <0$, where we have used
the unitarity of $\ttt$. On the other hand, since $\ttt$ is a
positive injective linear map, $\hat{\Psi}\equiv
\ttt(|\psi\rangle\langle\psi |)$ is a nonzero positive operator, so
that it admits a decomposition of the form $\hat{\Psi}=\sum_{k=1}^K
\epsilon_k\, |\phi_k\rangle\langle\phi_k |$, where
$\{\phi_k\}_{k=1}^K$ is an orthonormal system and
$\{\epsilon_k\}_{k=1}^K$ is a set of strictly positive numbers.
Therefore, since $\hat{B}$ is positive, we have:
\begin{equation}
\tr(\hat{\Psi}\,\hat{B})=\sum_{k=1}^K \epsilon_k\,
\tr(|\phi_k\rangle\langle\phi_k |\,\hat{B})=\sum_{k=1}^K
\epsilon_k\, \langle\phi_k, \hat{B}\,\phi_k\rangle >0.
\end{equation}
But this is in contrast with the inequality
$\tr(\hat{\Psi}\,\hat{B})<0$ previously found.~$\square$

From Lemma~{\ref{primo}} and Lemma~{\ref{secondo}} one obtains
immediately the following result:
\begin{proposition}
The inverse of a linear map belonging to the set $\oto\cap\ptm$
belongs to this set too. Hence, $\oto\cap\ptm$ is a subgroup of
$\oinv$.
\end{proposition}

We now consider the set $\pto$ consisting of those linear maps in
$\lh$ of the form $\ea\otimes\eb$, where $\ea$, $\eb$ are linear
maps in $\lha$, $\lhb$, respectively, that are positive,
trace-preserving and unitary. An analogous definition holds for the
set $\cpto$, with the `local maps' $\ea$, $\eb$ assumed to be
\emph{completely positive} rather than simply positive. It is clear
that the sets $\pto$ and $\cpto$ are semigroups with identity. We
will see that they are actually groups. Consider also the group of
\emph{local unitary transformations}
\begin{eqnarray} \label{def-otot}
\hspace{-0.5cm}\otot & := & \{\ttt\in\oto\colon\, \ttt(\hat{A}) =
(\hat{U} \otimes \hat{V})\hspace{0.3mm} \hat{A}\, (\hat{U}^\dagger
\otimes
\hat{V}^\dagger), \forall\hspace{0.3mm}\hat{A}\in\lh, \nonumber \\
& & \mbox{ for some unitary ops.\ $\hat{U},\hat{V}$ in $\lha,\lhb$
resp.} \},
\end{eqnarray}
which is obviously a subgroup of both $\pto$ and $\cpto$. In a
similar way one defines the group of \emph{local unitary-antiunitary
transformations} $\ototc$ (include local antiunitary operators
$\hat{U},\hat{V}$ in the r.h.s.\ of~{(\ref{def-otot})}). An example
of a map that belongs to $\ototc$, but not to $\otot$, is the tensor
product $\jja\otimes\jjb$ of two partial transpositions $\jja$ and
$\jjb$.
\begin{proposition}
The set $\pto$ is a group (with respect to composition of maps);
hence, it is a subgroup of the group $\oto\cap\ptm$.
\end{proposition}
\noindent {\bf Proof:} Given a map $\ea\otimes\eb$ in $\pto$
--- where $\ea$, $\eb$ are linear maps in $\lha$, $\lhb$,
respectively, that are positive, trace-preserving and unitary ---
the inverse map $\ea^{-1}\otimes\,\eb^{-1}$ is positive; to show
this, apply Lemma~{\ref{primo}} and Lemma~{\ref{secondo}}
identifying the (generic) Hilbert space $\mathcal{H}$ with the local
Hilbert spaces $\hilba$ and $\hilbb$.~{$\square$}

A \emph{Kadison automorphism} is a bijective map from $\densk$ ---
the convex set of density operators in a Hilbert space $\KK$ ---
onto itself that is convex linear. We now state as a lemma a
property of this kind of automorphisms that can be obtained `by
duality' from a well known result due to Kadison~\cite{Ktheorem}
(concerning $C^*$-algebras):
\begin{lemma}[Kadison] \label{Kadison-th}
Every Kadison automorphism $\ee\colon\densk\rightarrow\densk$ is of
the form
\begin{equation}
\ee (\hrho)=\hat{T} \hrho\, \hat{T}^\dagger,\ \ \
\forall\hrho\in\densk,
\end{equation}
where $\hat{T}$ is a unitary or anti-unitary operator.
\end{lemma}
Assume that the Hilbert space $\KK$ is finite-dimensional. Then,
since any operator $\hat{C}\in\lk$ can be written as
$\hat{C}=c_1\,\hrho_1-c_2\,\hrho_2$, for some
$\hrho_1,\hrho_2\in\densk$ and some non-negative numbers $c_1,c_2$,
it is clear that every Kadison automorphism
$\ee\colon\densk\rightarrow\densk$ extends (uniquely) in a natural
way to a linear map in $\lk$; conversely, a linear map in $\lk$
which is bijective on $\densk$ can be regarded as a Kadison
automorphism.

Let us denote by $\ot$ the group of unitary-antiunitary
transformations in $\lh$. We are now able to prove the following
result:
\begin{theorem}
The group $\oto\cap\ptm$ coincides with the group $\oto\cap\ot$. The
group $\pto$ coincides with the group $\ototc$. The set $\cpto$ is a
group which coincides with the group of local unitary
transformations $\otot$. All the mentioned groups are subgroups of
$\pts$, and $\pts$ is a subgroup of $\ot$.
\end{theorem}
\noindent {\bf Proof:} It is clear that the group $\oto\cap\ot$ is a
subgroup of $\oto\cap\ptm$. On the other hand, by
Lemma~{\ref{primo}} and Lemma~{\ref{secondo}}, a map in the group
$\oto\cap\ptm$ is a Kadison automorphism; hence, by
Lemma~{\ref{Kadison-th}}, it is contained in $\oto\cap\ot$. This
proves the first assertion of the theorem. Next, given a map
$\ea\otimes\eb$ in $\pto$
--- where $\ea$, $\eb$ are linear maps in $\lha$, $\lhb$,
respectively, that are positive, trace-preserving and unitary ---
the maps $\ea$ and $\eb$ are Kadison automorphisms. Hence, by
Lemma~{\ref{Kadison-th}},
\begin{equation} \label{unant}
\ea(\hat{A})=\hat{U} \hspace{0.3mm} \hat{A}\, \hat{U}^\dagger ,\
\forall\hspace{0.3mm}\hat{A}\in\lha,\ \ \ea(\hat{B})=\hat{V}
\hspace{0.3mm} \hat{B}\, \hat{V}^\dagger ,\
\forall\hspace{0.3mm}\hat{B}\in\lhb,
\end{equation}
for some unitary or antiunitary operators $\hat{U},\hat{V}$ in
$\lha,\lhb$, respectively. Therefore, the group $\pto$ coincides
with the group $\ototc$. Note that, since an antiunitary operator is
the composition of a unitary operator with a complex conjugation, if
in~{(\ref{unant})} we let the operators $\hat{U},\hat{V}$ be
antiunitary, we have that the maps $\ea$, $\eb$ are the composition
of unitary transformations with transpositions. Transpositions are
positive but not completely positive maps; hence, the set $\cpto$
coincides with the group of local unitary transformations $\otot$.
Finally, observe that the maps in the group $\pts$ are Kadison
automorphisms. Thus, by Lemma~{\ref{Kadison-th}}, the last assertion
of the theorem follows.~{$\square$}

\subsection{Local analysis} \label{local-analysis}

In~\cite{inter_pure}, the authors considered the Schmidt
decomposition of {\it pure} states, and, in that setting, analyzed
the geometry of the sets of Schmidt equivalent pure states (that
turn out to be differentiable manifolds). Since two pure states are
Schmidt equivalent if and only if they are mutually convertible via
local unitary transformations, they called such manifolds `the
manifolds of {\it interconvertible} states'. The aim of the present
section is to apply the same line of reasoning to study the
structure of the manifolds of Schmidt equivalent (mixed) states.
Here we assume for simplicity that $\na=\nb=N$, hence $\dd=N^2$.
This is not necessary for our purposes but it allows a convenient
simplification of our formulae.

As it is discussed below, there is a major limitation to the
straightforward extension of the aforementioned results from pure to
mixed states. The definition of SC's is based on the fact that
density operators are elements of a vector space. On the other hand,
they are constrained to be positive operators (of unit trace)
because of their physical interpretation. This leads to the main
difference between the results in~\cite{inter_pure} and our
forthcoming discussion: while in~\cite{inter_pure} the geometry of
the manifolds of interconvertible states can be characterized
globally, here we are limited to a `local' analysis (i.e.\ we are
forced to consider transformations in a neighborhood of the
identity).

As stated in Proposition~\ref{equivalence}, the transformations of
the form $ \hrho \mapsto {\hat{\mathcal T}}(\hrho), $ where
$\hat{\mathcal T}$ belongs to $\oto$, preserve the SC's. The
converse is also true: if two operators have the same SC's then they
are connected by a map in the group $\oto$. Therefore, we can build
the different equivalence classes acting locally on `fiducial
states' with the group $\oto$. Actually, we need to consider the
component connected to the identity of this group, which is
isomorphic to (and will be identified with) the Lie group
$\mathrm{SO}(\dd)\times\mathrm{SO}(\dd)$. We need then to impose two
conditions on the local (i.e.\ close to the identity)
transformations: {\it a)} the positivity of a fiducial state $\hrho$
must be preserved,  and {\it b)} the trace of $\hrho$ has to be
preserved as well. The first is a \emph{global constraint} that does
not `reduce the number of dimensions', although characterizing a
neighborhood of the identity in
$\mathrm{SO}(\dd)\times\mathrm{SO}(\dd)$ preserving the positivity
of a given fiducial state is a challenging open problem. The second
constraint, normalization, is a \emph{local constraint} that reduces
the dimension of the manifold by one: the condition that
$\tr(\hrho)=1$ amounts to fixing the projection of the vector
$\hrho$ along the direction $\id$. Let us consider an
orthonormal basis in the real Hilbert space of the form
$\{\id^{\sista}\otimes\id^{\sistb},\hat{F}^\sista_i\otimes\id^\sistb,
\id^\sista\otimes\hat{F}^\sistb_j,
\hat{F}^\sista_i\otimes\hat{F}^\sistb_j\}_{1\leq i,j\leq\dd-1}$;
clearly, $\tr(\hat{F}_i^\sista)=0=\tr(\hat{F}_j^\sistb)$, because of
the orthogonality condition with
$\id=\id^{\sista}\otimes\id^{\sistb}$. Given a state
$\hrho\in\dens$, we have:
\begin{equation}
\hrho=\frac{\id^\sista\otimes\id^\sistb}{\dd}+
\sum_{i=1}^{\dd-1}\alpha^\sista_i\hat{F}^\sista_i\otimes\id^\sistb+
\sum_{i=1}^{\dd-1}\alpha^\sistb_i\id^\sista\otimes\hat{F}^\sistb_i+
\sum_{i,j=1}^{\dd-1}\beta_{ij}\hat{F}^\sista_i\otimes\hat{F}^\sistb_j.
\end{equation}
The infinitesimal action of the Lie group
$\mathrm{SO}(\dd)\times\mathrm{SO}(\dd)$ on the basis elements is of
the form
\begin{eqnarray}
\hat{F}^{\sista,\sistb}_i & \mapsto & (\delta_{ij}+\phi^{\sista,\sistb}_{ij})\,\hat{F}^{\sista,\sistb}_j+
\epsilon^{\sista,\sistb}_i\; \id^{\sista,\sistb},\\
\id^{\sista,\sistb} & \mapsto &
\id^{\sista,\sistb}-\epsilon^{\sista,\sistb}_i\,\hat{F}^{\sista,\sistb}_i,
\end{eqnarray}
where contraction of repeated indices is understood.  Here
$\phi^\sista$ and $\phi^\sistb$ are real, $(\dd-1)\times(\dd-1)$
antisymmetric matrices and $\epsilon^\sista$ and $\epsilon^\sistb$
are real $(\dd-1)$-dimensional vectors.

\begin{figure}[htbp]
\centering
\includegraphics[height=0.35\textwidth]{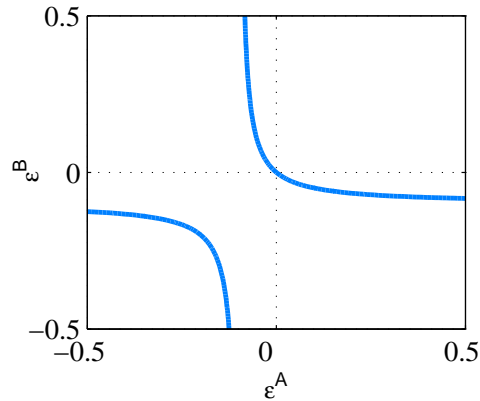}
\caption{An example of the parameter space of allowed orthogonal
transformations. Only the branch of the hyperboloid which contains
the origin must be considered.} \label{fig:doev}
\end{figure}

Hence, by applying an infinitesimal transformation in
$\mathrm{SO}(\dd)\times\mathrm{SO}(\dd)$ with generators
$(\phi^\sista,\epsilon^\sista;\phi^\sistb,\epsilon^\sistb)$ it is
not difficult to see that the coefficient of
$\id=\id^\sista\otimes\id^\sistb$ --- and hence $\tr(\hrho)$ --- undergoes the following change:
\begin{equation} \label{equa}
\delta\tr(\hrho)=\alpha^\sista_i\epsilon_i^\sista+\alpha^\sistb_i\epsilon_i^\sistb
+\beta_{ij}\epsilon_i^\sista\epsilon_j^\sistb.
\end{equation}
In this equation we have kept the leading non-trivial infinitesimal
changes, discarding those of the type $(\epsilon^{\sista})^2$ and
$(\epsilon^{\sistb})^2$. This is the correct expansion in the
space of jets of regular functions of two vectors with nonvanishing
gradients. Equation~{(\ref{equa})} defines a surface similar to an
hyperboloid. In order to visualize it one can simply take
${\epsilon_i^{\sista,\sistb}}=\delta_{1i}\epsilon^{\sista,\sistb}$.
For a simple choice of $\alpha,\beta$, the result is plotted in
figure \ref{fig:doev}. Clearly, connection to the identity
implies that only the branch containing the origin
has to be considered.

Therefore, the vectors $\epsilon^{\sista,\sistb}$ are constrained
while the matrices $\phi^{\sista,\sistb}$ are
unconstrained.\footnote{We are neglecting here the case
$\alpha_i^{\sista,\sistb}=0$, $\beta_{ij}\neq 0$. In this atypical
case we need to keep terms of $\Ord{\epsilon^2}$ in the
transformation law of $\id^{\sista,\sistb}$. Once this is done one
checks that the codimension is still one. The really degenerate case
$\alpha=\beta=0$, instead, corresponds to the class of equivalence
of a single point $\hrho=\id^\sista\otimes\id^\sistb/\dd$, with
dimension 0. This point has to be considered as the `tip', the
extremal point of the space of states.} Assuming that {\it the SC's
are all different} (i.e.\ that $\hrho$ is a `typical state'), one
finds out that the Schmidt equivalence class $\scrho$ is locally
diffeomorphic to
\begin{equation}
\frac{\mathrm{SO}(\dd)\times\mathrm{SO}(\dd)}{\mathrm{SO}(2)},
\end{equation}
where the subgroup $\mathrm{SO}(2)$ is generated by the linear
combination of the coefficients
$\{\epsilon_i^\sista,\epsilon_i^\sistb\}_{i=1}^{\dd-1}$ derived
above. Then, the dimension of the manifold $\scrho$ is
\begin{equation}
\mathrm{d}_{\hrho}=2\frac{\dd(\dd-1)}{2}-1=\dd^2-\dd-1.
\end{equation}
It is easy to see how this counting is consistent with the intuition
that changing one of the SC's $\lambda_1,\ldots,\lambda_{\dd}$ of $\hrho$ brings out of
the equivalence class $\scrho$. Indeed, subtracting the number of the SC's from
the number of dimensions --- $(\dd^2-1)-\dd=\dd^2-\dd-1$ --- we obtain the
dimensionality of a typical orbit. For example, take the case of
two-qubits: $\na=\nb=2$. We have: $\dd=4$ and  $\dd^2-1=15$; hence,
the manifold $\scrho$, for a typical bipartite state $\hrho\in\dens$, is $(15-4=11)$-dimensional.

If $\hrho\in\dens$ is a non-typical state, the dimension of the manifold $\scrho$ is smaller than
$\dd^2-\dd-1$. In fact, in the case where the $\dd$ SC's of $\hrho$
cluster into subsets of $m_1,...,m_h$
identical values --- by means of an argument analogous to the one adopted in~\cite{inter_pure}
--- one can check that there is a `local stabilizer subgroup' isomorphic to
$\mathrm{SO}(m_1)\times \dots \times \mathrm{SO}(m_h)$; therefore, in this case,
$\scrho$ is locally diffeomorphic to
\begin{equation}
\frac{\mathrm{SO}(\dd)\times\mathrm{SO}(\dd)/\mathrm{SO}(2)}{
\mathrm{SO}(m_1)\times \dots \times \mathrm{SO}(m_h)}.
\end{equation}
Then, the dimension $\mathrm{d}_{\hrho}$ of $\scrho$ is given, in
general, by
\begin{equation}
\mathrm{d}_{\hrho}=\dd(\dd-1)-1-\sum_{k=1}^h\frac{m_k(m_k-1)}{2}.
\end{equation}

In order to make the above argument \emph{rigorous}, one should
actually \emph{prove} that the Schmidt equivalence classes are
actually differentiable manifolds. In that case our previous
argument would provide the correct dimension of such manifolds. As
we learn from mathematicians~\cite{Raja}, a standard tool for
characterizing a subset of a differentiable manifold (like the
manifold of Hermitian operators in $\mathcal{H}$) as a submanifold
is the study of the (possible) Lie groups acting transitively on the
given subset and of the associated stabilizer subgroups. Therefore,
suitably improving the analysis of subsection~{\ref{local_maps}} may
allow to achieve such a remarkable result.

\section{\label{ideas}Entanglement and symmetric polynomials in the Schmidt coefficients}

We will now consider the role played by Schmidt coefficients in the characterization of
entanglement. Our starting point will be the realignment criterion (RC). It is natural to
wonder if the \emph{whole set} of the SC's of a bipartite state may allow a stronger characterization
of entanglement with respect to the RC.

As we have seen, the RC, like other separability criteria, is based
on the evaluation of a single functional in terms of which a
necessary condition for separability can be stated. In the case of
the RC, that functional coincides with the sum of the Schmidt
coefficients. On the other hand the evaluation of a single
functional might not be sufficient to determine completely the
presence of entanglement \cite{more_than_one} (from a more general
point of view, we may say, to characterize classical and quantum
correlations). In particular, as the RC is only a necessary
condition for separability, one is led to consider additional
functionals in order to gain information about the presence of
entanglement.

Here we propose to consider the {\it symmetric polynomials} in the
Schmidt coefficients, namely:
\begin{eqnarray}
\begin{array}{ccl}
\minor^{[1]} & = & \sum_{k=1}^\dd \lambda_k \\
\minor^{[2]} & = & \sum_{h\neq k=1}^\dd \lambda_h \lambda_k \\
& \dots & \\
\minor^{[l]} & = &\sum_{\{ i_1,i_2\dots i_l \}} \lambda_{i_1}
\lambda_{i_2} \dots \lambda_{i_l}\\
&  \dots & \\
\minor^{[\dd]} & = & \prod_{k=1}^\dd \lambda_k.
\end{array}
\end{eqnarray}
Notice that the RC involves the symmetric polynomial of degree one.

A naive argument says that, if the sum of the Schmidt Coefficients
is equal to $S$, their product is upper bounded by $(S/\dd)^{\dd}$.
Hence we have the following condition for a separable density matrix
$\rho$:
\begin{equation}\label{Det}
\hrho \ \ \mbox{separable} \ \ \Rightarrow \ \ \minor^{[\dd]} \leq
\left( \frac{1}{\dd} \right)^\dd,
\end{equation}
which obviously defines a weaker separability criterion.
Analogously, one can consider the symmetric polynomial of degree $l$
and obtain the following necessary conditions for separability:
\begin{equation}\label{naive}
\hrho \ \ \mbox{separable} \ \ \Rightarrow \ \ \minor^{[l]} \leq
y_l(\dd) = {\dd \choose l} \left( \frac{1}{\dd} \right)^{l}.
\end{equation}
In particular, if $\hrho$ has Schmidt rank $\RR$, we can write the
conditions
\begin{equation}\label{Minors}
\minor^{[l]} \leq {\RR \choose l} \left( \frac{1}{\RR} \right)^l
\end{equation}
for $l \leq \RR$, while $\minor^{[l]}=0$ for $l>\RR$.

It is worth noticing that the symmetric polynomials
$\{\minor^{[l]}\}$ are in one-to-one correspondence with the Schmidt
coefficients $\{ \lambda_k \}$. The inequalities (\ref{naive}) are
consequences of the RC. Hence, as separability criteria, they are
weaker than the parent one. Section \ref{new} will be devoted to the
study of possible stronger generalization.

Finally, we notice that in the approach followed in \cite{ReCr},
which makes use of the associated realigned matrix $\rho^\rea$, the
symmetric polynomials are the coefficients of the characteristic
polynomial:
\begin{equation}\label{c.poly}
\chi_{\rho^\rea}(x) = \det (|\rho^\rea|-x\mathbbm{I}) =
\sum_{l=1}^\dd \minor^{[l]}(|\rho^\rea|) (-x)^{\dd-l}
\end{equation}
where, with abuse of notation, we indicated with $\minor^{[l]}(A)$
the principal minor of order $l$ of the matrix $A$.

\section{\label{duality}Quantum states and quantum maps}

This section is devoted to the application of the ideas presented in
section \ref{ideas} to the study of quantum channels, i.e.\
completely positive trace-preserving (CPT) maps. In order to do
that, we exploit the well known correspondence between quantum
channels and quantum states (see \cite{MaSt,Jam,ZycBen}).

Here we consider quantum systems with Hilbert spaces $\hilba$ and
$\hilbb$, the set of states in the composite system
$\mathcal{D}(\hilba\otimes\hilbb)$, and the set of CPT maps from
system $\mathsf{B}$ to system $\mathsf{A}$, which is denoted
$\mathrm{CPT}(\hilbb,\hilba)$.

Given a CPT map $\hat\mathcal{E} \in \mathrm{CPT}(\hilbb,\hilba)$,
one can associate a state $\hat\rho \in
\mathcal{D}(\hilba\otimes\hilbb)$ in the following {\it canonical}
way:
\begin{equation}\label{can}
\hat\mathcal{E} \ \ \longrightarrow \ \ \hat\rho = ( \hat\mathcal{E}
\otimes \hat\mathcal{I} ) (\hat\beta),
\end{equation}
where $\hat\beta = |\phi\rangle\langle\phi| \in
\mathcal{D}(\hilbb\otimes\hilbb)$ denotes a maximally entangled
state, for instance $|\phi\rangle = \frac{1}{\sqrt{\nb}}
\sum_{\alpha=1}^\nb |\alpha\rangle|\alpha\rangle$, and
$\hat\mathcal{I}$ is the identical map in the system $\mathsf{B}$.

A CPT map $\hat\mathcal{E}$ is said to be entanglement breaking (EB)
if $\hat\mathcal{E}\otimes\hat\mathcal{I}$ maps any state into a
separable one \cite{EnBr}.
One can show that a CPT is EB if and only if the map
$\hat\mathcal{E}\otimes\hat\mathcal{I}$ transforms a maximally
entangled state into a separable one.
It follows that the CPT map is EB if and only if the canonically
associated state is separable.

One can select a local orthogonal basis $\hat F_{(m n)}^\mathsf{A} =
|m\rangle\langle n|$ and $\hat F_{(\mu \nu)}^\mathsf{B} =
|\mu\rangle\langle\nu|$ and write the matrix elements of the CPT map
in that basis as follows
\begin{equation}
\mathcal{E}_{(mn)(\mu\nu)} = \tr\left( \hat F_{(m n)}^\mathsf{A}
\hat\mathcal{E}\left(\hat F_{(\mu \nu)}^\mathsf{B}\right) \right).
\end{equation}

It is easy to check that, with the canonical association
(\ref{can}), the matrix representation of the state $\hrho$ and the
map $\hat\mathcal{E}$ are related in the following way:
\begin{equation}
\rho_{(i\alpha)(j\beta)} = \frac{1}{\nb}
\mathcal{E}_{(ij)(\alpha\beta)}.
\end{equation}

Following the definition in \cite{ReCr}, it is immediate to
recognize that, apart of the normalization factor $1/\nb$, the
matrix expression of $\hat\mathcal{E}$ is identical to the realigned
matrix $\rho^\rea$ (see equation (\ref{rea_matrix})).
Identifying the CPT map with the realigned matrix of the
corresponding density matrix, one can consider the characteristic
polynomial
\begin{equation}
P_{\mathcal{E}}(x) = \det ( |\mathcal{E}| - x \mathbbm{I} ),
\end{equation}
where $|\mathcal{E}| = \sqrt{\mathcal{E}^\dag \mathcal{E}}$ and
$\mathcal{E} \equiv \mathcal{E}_{(ij)(\alpha\beta)}$.
To fix the ideas, let us consider the case in which $\mathcal{E}$
has full rank. One obtains from (\ref{Det}) the following necessary
condition for $\hat\mathcal{E}$ to be entanglement breaking:
\begin{equation}
\det\left(|\mathcal{E}|\right)\leq\left(\frac{\nb}{\dd}\right)^\dd.
\end{equation}

The determinant of quantum channels was also considered
in~\cite{Wolf}, in which some of its properties were presented and
discussed in the context of factorization of CPT maps. The present
result relates a geometric property of the map, such as the rate of
contraction of volume (which is equal to
$\det\left(|\mathcal{E}|\right)$) to the property of being
entanglement breaking.
Analogously, from (\ref{Minors}), if the matrix
$\mathcal{E}_{(ij)(\alpha\beta)}$ has rank $\RR$, we can write the
following necessary conditions for $\hat\mathcal{E}$ to be
entanglement breaking:
\begin{equation}
\minor^{[l]}\left(|\mathcal{E}|\right) \leq {\RR \choose l}
\left(\frac{\nb}{\RR} \right)^{l}
\end{equation}
for $l=1\dots \RR$.

\section{\label{new}Beyond the realignment criterion}

In section~\ref{ideas} we introduced a family of separability
conditions which are weaker than (or equivalent to) the RC.
In this section we argue about the possibility of extending the
family of separability criteria defined in (\ref{naive}) in order to
write criteria which are independent of the RC.
As a first step in this direction, we may ask whether it is possible
to find strict upper bounds $x_l(\dd,\DD)$ such that:
\begin{equation}\label{strict}
\hrho \ \ \mbox{separable} \ \ \Rightarrow \ \ \minor^{[l]} \leq
x_l(\dd,\DD) < {\dd \choose l} \left( \frac{1}{\dd} \right)^{l}.
\end{equation}
However, in the following we consider a weaker
statement\footnote{Notice that (\ref{LowerBound}) is weaker as
separability criterion, while it is stronger in the sense that
(\ref{LowerBound}) implies (\ref{strict}). In particular
$x_l(\dd,\DD) \leq \tilde{x}_l(\dd,\DD)$}, namely
\begin{equation}\label{LowerBound}
\minor^{[1]} \leq 1 \ \ \Rightarrow \ \ \minor^{[l]} \leq
\tilde{x}_l(\dd,\DD) < {\dd \choose l} \left( \frac{1}{\dd}
\right)^{l}
\end{equation}
which establishes a strict upper bound for the functionals
$\minor^{[l]}$ over the set of states satisfying the RC.

The following proposition holds true:

\begin{proposition}\label{XXX}
The upper bounds $\tilde{x}_l(\dd,\DD)$ in equation
(\ref{LowerBound}) exist for $\DD < \dd^3$.
\end{proposition}

\noindent {\bf Proof:} The proposition is proven by contradiction.
Let us suppose the existence of a density matrix $\rho_0$ such that
$\minor^{[1]} \leq 1$ and the inequalities in (\ref{naive}) are
saturated. That implies that the density matrix has maximum rank,
$\RR = \dd$, and all its SC's are all equal to $\dd^{-1}$. Hence,
referring to the singular value decomposition in equation
(\ref{ref_me}), we can write the Schmidt decomposition of the
density matrix in the following way:
\begin{equation}\label{proof_Sch}
{\rho_0}_{(\alpha i)(\beta j)} = \frac{1}{\dd} \sum_{k,l=1}^\dd
u_{(\alpha\beta)(kl)} {v^*}_{(kl)(ij)},
\end{equation}
where $v_{(ij)(i'j')}$ and $u_{(\alpha\beta)(\alpha'\beta')}$ are
respectively the entries of the unitary matrices $\mathcal{V}$ and
$\mathcal{U}$ (see equation (\ref{ref_me})), with dimension $\na^2$
and $\nb^2$.
We introduce the following notation
\begin{equation}
u_{(\alpha\beta)(kl)} {v^*}_{(kl)(ij)} = \langle
\hat{v}_{ij},\hat{u}_{\alpha\beta} \rangle,
\end{equation}
where $\hat{u}_{\alpha\beta}$ and $\hat{v}_{ij}$ indicate the
vectors respectively defined as the rows and the columns of the
matrices $\mathcal{U}$ and $\mathcal{V}$.
We have:
\begin{equation}
\tr\left(\rho_0\right) = \frac{1}{\dd^2} \sum_{i,\alpha} \langle
\hat{v}_{ii},\hat{u}_{\alpha\alpha} \rangle = \frac{1}{\dd^2}
\langle \sum_{i=1}^{\na} \hat{v}_{ii},\sum_{\alpha=1}^{\nb}
\hat{u}_{\alpha\alpha}\rangle,
\end{equation}
and we obtain:
\begin{equation}\label{contrad}
\tr\left(\rho_0\right) = |\tr\left(\rho_0\right)| = \frac{|\langle
\sum_i \hat{v}_{ii} , \sum_\alpha
\hat{u}_{\alpha\alpha}\rangle|}{\dd^2} \leq \frac{|\sum_i
\hat{v}_{ii}| |\sum_\alpha \hat{u}_{\alpha\alpha}|}{\dd^2} =
\frac{\sqrt{\na\nb}}{\dd^2},
\end{equation}
(the last equality holds true since $\hat{u}_{\alpha\alpha}$ and
$\hat{v}_{ii}$ are two systems of orthonormal vectors) which, for
$\DD < \dd^3$, is in contradiction with the hypothesis that $\rho_0$
has unit trace.
Since the set of states satisfying the RC is compact and the
symmetric polynomials $\minor^{[l]}$ are continuous functionals, the
lower upper bounds $\tilde{x}_l(\dd,\DD)$ do exist. $\Box$

\begin{figure}
\centering
\includegraphics[height=0.32\textwidth]{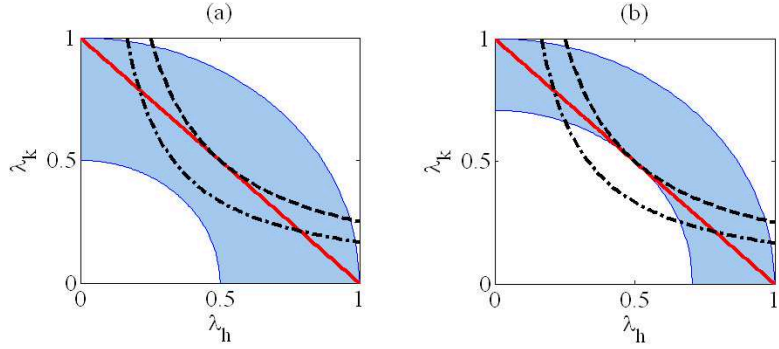}
\caption{A pictorial representation of the set of the equivalence
classes, for the generic case $\dd \neq \DD$ (a); and for the
degenerate one $\dd = \DD$ (b). The Schmidt coefficients are
represented along the axes. States are contained in the colored
region, which fulfills the constraints on the purity (\ref{purity}).
The region on the bottom-left of the solid line fulfill the RC
(hence separable states are contained in this region). The region on
the bottom-left of the dashed line contains the states that fulfill
one of the {\it naive} inequalities (\ref{naive}) for some $l>1$.
Finally, the region on the bottom-left of the dot-dashed line
contains the states that fulfill the proposed criterion RC$l$.
Notice that the RC (or RC$1$) bounds separable states with a
hyperplane, while the higher order RC$l$, for $l>1$, bounds the set
of separable states with non-linear hypersurfaces. Notice that the
RC$l$'s can in principle allow a finer characterization of entangled
states.} \label{bound}
\end{figure}

In the following we indicate with RC$l$ the suggested criterion
\begin{equation}
\hrho \ \ \mbox{separable} \ \ \Rightarrow \ \ \minor^{[l]} \leq
x_l(\dd,\DD).
\end{equation}
Figure \ref{bound} shows a pictorial representation of the relation
between the parent RC, the weaker criteria (\ref{naive}) and the
proposed extensions RC$l$.
As the figure suggests, the RC$l$ can in principle be used, with
respect to the information given by the RC, as refinements of the
knowledge about the region of separable states.

It is worth noticing, however, that there are still two main open
problems:
\begin{enumerate}
\item
the actual values of the upper bounds $x_l(\dd,\DD)$ (as well as
$\tilde{x}_l(\dd,\DD)$) are still undetermined;

\item
it is not clear whether the criteria RC$l$ are independent of the
RC, i.e.\ if there are entangled states such that RC is not violated
while RC$l$ is for some $l>1$. That is equivalent to the strict
inequality $x_l(\dd,\DD) < \tilde{x}_l(\dd,\DD)$.
\end{enumerate}
In the next section we face these problems with a numerical
approach. We are going to restrict our discussion to the case of
lower dimensional systems, namely $\hilba\otimes\hilbb =
\mathbbm{C}^2\otimes\mathbbm{C}^2,
\mathbbm{C}^2\otimes\mathbbm{C}^3$. For this cases one can exploit
the fact that the PPT (positive partial transpose) criterion
\cite{Horo96} is necessary and sufficient for separability.

\section{\label{examples}Examples for low dimensional systems}

In analogy to what can be done for the RC (see for instance
\cite{Aniello}), one could determine the value of the strict upper
bounds $x_l(\dd,\DD)$ by convex linearity starting from the
properties of pure separable states. Nevertheless, it is worth
noticing that that can be a rather difficult task since the
symmetric polynomials $\minor^{[l]}$ are not easy to manipulate with
respect to the convex structure of the set of separable states. For
this reason, in the following we present a numerical analysis which
allows to present some interesting results.

For a preliminary analysis of the potentialities of the proposed
family of criteria, they have been numerically tested in the case of
a bipartite qubit-qubit and qubit-qutrit system. A numerical search
of the upper bounds can be done exploiting the PPT criterion
\cite{Horo96}.
The constraints $\rho\geq 0$ and $\rho^{T_A}\geq 0$, where $T_A$
indicates the partial transposition, are known to be necessary and
sufficient to characterize separable states in the low dimensional
cases. Hence, the determination of the lower upper bounds
$x_l(\dd,\DD)$ reduces to a problem of constrained maximization.

\begin{figure}
\centering
\includegraphics[height=0.32\textwidth]{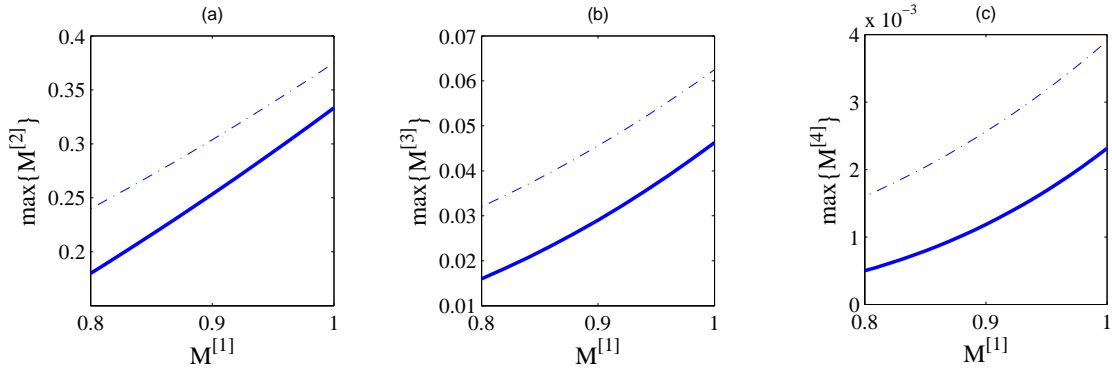}
\caption{Numerically estimated upper bounds of the symmetric
polynomials $\minor^{[2]}$ (a), $\minor^{[3]}$ (b), and
$\minor^{[4]}$ (c), as functions of the value of $\minor^{[1]}$ for
a qubit-qubit system: naive upper bounds (dash-dotted line),
numerically estimated upper bounds over the set of separable states
(solid line) which coincides with upper bounds over the set of all
states (separable and entangled).} \label{2x2Ml}
\end{figure}

\begin{figure}
\centering
\includegraphics[height=0.32\textwidth]{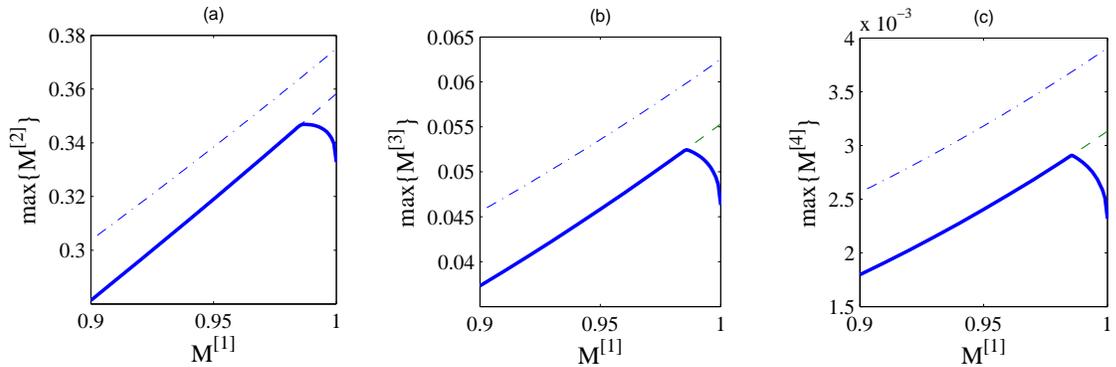}
\caption{Numerically estimated upper bounds of the symmetric
polynomials $\minor^{[2]}$ (a), $\minor^{[3]}$ (b), and
$\minor^{[4]}$ (c), as functions of the value of $\minor^{[1]}$ for
a qubit-qutrit system: naive upper bounds (dash-dotted line),
numerically estimated upper bounds over the set of separable states
(solid line) and over the set of all states (dashed line). Notice
the presence of a region in which the functionals have a lower upper
bound over the set of separable states.} \label{2x3Ml}
\end{figure}

We have numerically estimated the maxima of the functions
$\minor^{[l]}$ over separable states (hence determining estimates
for $x_l(\dd,\DD)$), and over the set of states satisfying the RC
(hence estimating the upper bounds $\tilde{x}_l(\dd,\DD)$). For a
qubit-qubit system ($\dd=\DD=4$) the results are shown in table
\ref{RC_2X2} together with {\it naive} bounds in (\ref{naive}). The
analogous quantities are shown in table \ref{RC_2X3} for the case of
a qubit-qutrit system ($\dd=4$, $\DD=9$). In the latter case we
found $x_l(\dd,\DD) < \tilde{x}_l(\dd,\DD)$ suggesting that the
criteria RC$l$ {\it can be in principle} stronger than the RC.
\begin{table}[hbtp]
\begin{center}
\begin{tabular}{|p{1.5cm}||*{4}{p{1.5cm}|}} \hline
                   & $l=1$ & $l=2$    & $l=3$     & $l=4$      \\
\hline \hline
$y_l(4)$           & $1$   & $0.3750$ & $0.06250$ & $0.003906$ \\
\hline
$\tilde{x}_l(4,4)$ & $1$   & $0.3333$ & $0.04630$ & $0.00231$ \\
\hline
$x_l(4,4)$         & $1$   & $0.3333$ & $0.04630$ & $0.00231$  \\
\hline
\end{tabular}
\end{center}
\caption{\label{RC_2X2}For a qubit-qubit system, the table shows the
upper bounds on the sums of the symmetric polynomials
$\minor^{[l]}$. $y_l(4)$ denotes the naive bounds (\ref{naive}).
$\tilde{x}_l(4,4)$ denotes the numerically estimated strict bounds
over the states satisfying the RC (\ref{LowerBound}). Finally,
$x_l(4,4)$ indicates the numerically estimated strict bounds over
the set of separable states (\ref{strict}). Notice that in this case
we found $x_l(\dd,\DD) = \tilde{x}_l(\dd,\DD)$ suggesting that the
criteria RC$l$ {\it cannot be} independent of the RC.}
\end{table}
\begin{table}[hpbt]
\begin{center}
\begin{tabular}{|p{1.5cm}||*{4}{p{1.5cm}|}} \hline
                   & $l=1$ & $l=2$    & $l=3$     & $l=4$      \\
\hline \hline
$y_l(4)$           & $1$   & $0.3750$ & $0.06250$ & $0.003906$ \\
\hline
$\tilde{x}_l(4,9)$ & $1$   & $0.3583$ & $0.05533$ & $0.003133$ \\
\hline
$x_l(4,9)$         & $1$   & $0.3469$ & $0.05249$ & $0.00291$  \\
\hline
\end{tabular}
\end{center}
\caption{\label{RC_2X3}For a qubit-qutrit system, the table shows
the upper bounds on the sums of the symmetric polynomials
$\minor^{[l]}$. $y_l(4)$ denotes the naive bounds (\ref{naive}).
$\tilde{x}_l(4,9)$ denotes the numerically estimated strict bounds
over the states satisfying the RC (\ref{LowerBound}). Finally,
$x_l(4,9)$ indicates the numerically estimated strict bounds over
the set of separable states (\ref{strict}). Notice that in this case
we found $x_l(\dd,\DD) < \tilde{x}_l(\dd,\DD)$ suggesting that the
criteria RC$l$ {\it can be} in principle independent of the RC.}
\end{table}
Figures \ref{2x2Ml} and \ref{2x3Ml} show the maximum of the
functionals $\minor^{[l]}$ for $l=2,3,4$, computed for fixed values
of $\minor^{[1]}$, respectively for qubit-qubit and qubit-qutrit
system, as functions of the value of $\minor^{[1]}$. The maxima are
computed over generic states and over separable states. In the case
of qubit-qutrit system, the plots show the region in which the
criteria RC$l$ can in principle be stronger than
--- or independent to --- the RC.

To conclude this section, we consider the case of two-qubit
(generalized) Werner states, of the form:
\begin{equation}\label{werner}
\hrho_p = p |\phi\rangle\langle\phi| + (1-p) \hat\mathbbm{I}/4 \;,
\end{equation}
for $p \in [0,1]$, where $|\phi\rangle$ indicates a maximally
entangled pure state.
The Schmidt coefficients of the state (\ref{werner}) are easily
calculated to be $\{ 1/2, p/2, p/2, p/2 \}$, yielding
\begin{eqnarray}
\begin{array}{ccl}
\minor^{[1]} & = & (1 + 3p)/2 \\
\minor^{[2]} & = & 3(p + p^2)/4 \\
\minor^{[3]} & = & (3p^2 + p^3)/8\\
\minor^{[4]} & = & p^3/16.
\end{array}
\end{eqnarray}
Notice that the symmetric polynomials are monotonically increasing
functions of the state parameter $p$.
It was shown in \cite{Rudolph} that the realignment criterion is
necessary and sufficient for this family of states (indeed it is so
for all the two-qubit states with {\it maximally disordered
subsystems}).
The state in (\ref{werner}) is known to be separable for $p \in
[0,1/3]$ and entangled otherwise.
We can compute the maximal value of the symmetric polynomials
$\mathrm{M}^{[l]}$ over the separable states in that family. These
maxima are reached in correspondence of the value $p=1/3$, hence
yielding:
\begin{eqnarray}\label{wer}
\hrho_p \ \ \mbox{separable} \ \ \Rightarrow \ \
\begin{array}{ccl}
\minor^{[1]} & \leq & 1 \\
\minor^{[2]} & \leq & 1/3 \\
\minor^{[3]} & \leq & 5/108 \simeq 0.04630 \\
\minor^{[4]} & \leq & 1/432 \simeq 0.00231.
\end{array}
\end{eqnarray}
Notice that the values in (\ref{wer}) computed for two-qubit
generalized Werner states saturates the numerical estimated upper
bounds reported in the table \ref{RC_2X2}.

\section{\label{econcludo}Conclusions}

The main goal of the present paper is to bring attention to the
Schmidt coefficients of a bipartite density operator, and to their
role for entanglement detection. The notion of Schmidt equivalence
classes has been introduced and a preliminary characterization of
such classes has been provided.

We have presented a family of separability criteria, written in
terms of the Schmidt coefficients, which are derived from the
realignment criterion. These separability criteria are consequence
of the fact that the symmetric polynomials in the Schmidt
coefficients are upper bounded on the set of separable states.

The application of that family of criteria to the study of quantum
channels determines a relation between a physical feature, such as
the preservation of entanglement under the action of the channel,
and a geometrical quantity, such as the determinant --- or the sum
of principal minors of order $l$ --- of a corresponding matrix.

We conjecture --- also with support of numerical examples
--- that a strengthened version of these criteria, independent of
the realignment criterion, exists. In particular, we have given
numerical examples for the case of the qubit-qutrit system. These
numerical results are of course not sufficient for achieving
independent separability criteria. However, they can open the way to
an analytical determination of stricter upper bounds on the
symmetric polynomials, and this may eventually lead to new
separability criteria.

\ack The authors wish to thank G.~Marmo for invaluable human and
scientific support, and F.~Lizzi for suggestions and encouragements.
C.~L.\ thanks M.~M.~Wolf and J.~I.~Cirac for suggestions and
discussions which led to the realization of the present work. C.~L.\
acknowledges the support of the project CONQUEST,
MRTN-CT-2003-505089. The main results of the paper were presented by
one of the authors (C.~L.) at the international conference \emph{The
Jubilee 40th Symposium on Mathematical Physics -- Geometry \&
Quanta} (25-28 June 2008, Torun, Poland). He wishes to thank the
organizers for their very kind hospitality.

\end{document}